# Severity Prediction of Drought in A Large Geographical Area Using Distributed Wireless Sensor Networks

Satish.G. Dappin,
*Department of Computer Science and Engineering, Dayananda Sagar College of Engineering, Bangalore, India. Email:* satishdappin@gmail.com

Vaidehi M,
*Research Associate, Research and Industry Incubation Centre, Dayananda Sagar College of Engineering, Bangalore, India. Email:*vaidehidm@yahoo.co.in

Nithya G. Nair,
*Department of Electrical and Computer Engineering, Illinois Institute of Technology, Chicago, IL60616,USA. Email:* nithya_g_nair@yahoo.com

Dr. T.R.Gopalakrsihnan Nair,
*Director, Research and Industry, RIIC, Professor, SM IEEE Dayananda Sagar College of Engineering, Bangalore, India. Email: trgnair@yahoo.com*

***Abstract***

**In this paper, the severity prediction of drought through the implementation of modern sensor networks is discussed. We describe how to design a drought prediction system using wireless sensor networks. This paper will describe a terrestrial interconnected wireless sensor network paradigm for the prediction of severity of drought over a vast area of 10,000 sq km. The communication architecture for sensor network is outlined and the protocols developed for each layer is explored. The data integration model and sensor data analysis at the central computer is explained. The advantages and limitations are discussed along with the use of wireless standards. They are analyzed for its relevance. Finally a conclusion is presented along with open research issues.**

## 1. Introduction

Wireless sensor network has the potential to solve the most of the environmental problems such as natural disaster like drought. Application like drought prediction depends on the measurement of the environmental factors that are based on the availability of sensor data. A large amount of gathered data from sensors of different locations and different points with real time processing leads to more acute prediction of the drought. Severity of drought prediction using the traditional methods is complicated by the variant phenomenon of natural parameters. Efficient drought prediction is not possible in traditional methods as they lack the real time information from the system [2]. With wireless sensor networks timely information on drought effecting parameters is provided to the decision makers and to the users. Wireless sensor network can cover large area, allowing application to access data from large amounts of sensors in real time; data can be aggregated and processed within the network. This helps to be successful in detecting onset, end and the accumulated stress of drought.

## 2. Wireless sensor network for the severity prediction of drought

In our drought severity prediction application sensor nodes collect measurement data such as temperature, humidity, wind pressure, precipitation and Wireless sensor network has the ability to communicate between the peer nodes and to the base station. Sensors nodes are deployed densely in large geographical area, which has the capacity to assemble together, establish routing topology and transmit data to the collection point. Our application highly requires the network to be robust, scalable and fault tolerant. Since the monitoring of the data is needed for long time of the year, network should run with low power consumption, engineers should have the capability to iterate the nodes as and when required. System should cost less and must be easily deployable. The human involvement is minimized by the use of wireless sensor networks in the whole operations.

## 3. Related Work

The network is divided into five sub sensor networks and nodes are deployed in the sensor networks communicating with the local base station. Sensor nodes are embedded with analog sensors like temperature, humidity, wind pressure, wind direction, wind pressure and precipitation on the sensor board [1]. These sensors senses the stimulus from the environment and analog data will be given to analog to digital converter of the computational part of the node called mote.

This will contain the microcontroller as a computational device. Radio on the mote is used to communicate with peer nodes with radio frequency communication. RF communication forms the sensor network at the local site. The sensed data will be transmitted to the base station through the nodes in the path. The sensor node contains message processing unit which will pass the messages sent by neighboring node to its immediate next node and sensing unit senses the changes in the environmental conditions and message is sent out through message processing unit and Base station at the local site consists of memory for the database storage. Memory varies with the base stations from different manufacturers.

The entire base stations are connected through wireless LANs. The range of wireless LAN can be extended to 80-120 kms by using high gain 802.11 networks with high gain antennas connected in line of sight. TCP/IP protocols are used in the 802.11 networks. Network costs less and the users will have more control over it. With 802.11 networks each node will have its own IP address and can be accessed or controlled over Internet from a remote base station. The use of intelligent routing protocols will enable sensors to form network and thereby enabling routing of the information in the absence of direct line of sight to the base station. Ultimate aim of the sensor network is accomplished by transferring the data from all the base stations of local site to the remote base station, which will consist of high-speed cluster computers. Experts can later manipulate the data. On this platform we run the draught detection, arrival probability and severity prediction applications and the results can be made available on real-time basis to the various users like the decision makers and meteorologists for drought pattern evolution.

## 4. System architecture

The proposed system architecture of covering 10000sq km is based on 3-tired architecture. Tier 1 consists of sensor nodes at the local RF site, which perform sensing the parameters like temperature, wind speed, wind direction, wind pressure, ground water presence and precipitation. Computational part does general purpose computing of the sensed parameters. Large geographical area is divided into 5 chunks of sensor networks each sensor network contains sensor nodes that roughly cover 100 sq km area. Which is a small trial network that is deployed first then nodes are added towards the area of phenomenon of interest as shown in “Fig.1,”. Sensor nodes are densely deployed in pre-identified region of the large geographical area of consideration. Each region is widely separated by each other. Each sensor node collects data of its surroundings as they are placed at the phenomenon of interest. Special high resolution can be achieved by deploying sensor nodes densely in the area. Motes will provide the computational module, which is a programmable unit that provides the computation, storage and bi-directional communication with the neighboring nodes of the network. Each individual sensor nodes communicate and coordinate

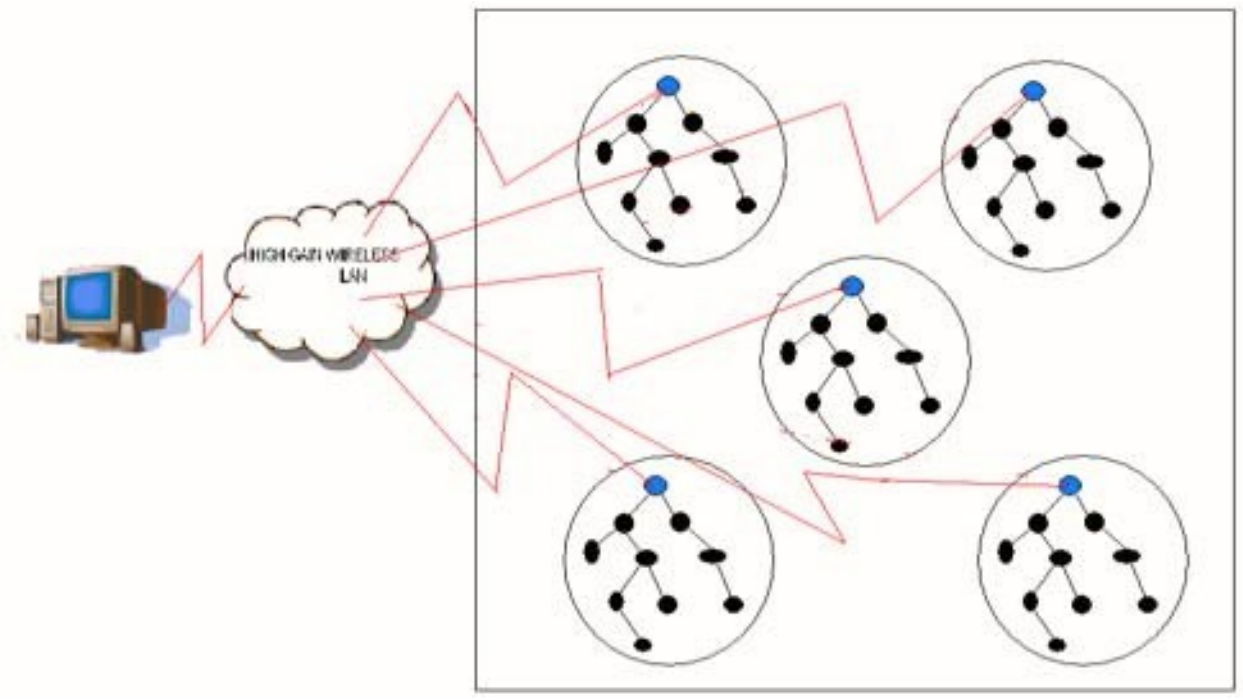

Fig.1.System architecture

themselves, forms multihop network by forwarding the data to the local base station.

All the local base stations connected each other in line of sight forms tier 2 of the architecture. All the base stations are connected to the remote base station. All the local base station will transfer the collected data to the remote base station which is further sent to high speed cluster computers for data manipulation. We want base station to perform reliable without fail, and since they consumes a lot of power, they are provided with adequate power capabilities and housing capability. All the base stations are connected with high gain wireless LAN capabilities to remote base station. Local base stations are connected with the remote base station and central processing unit with wide area connectivity. Remote end users can access the data with WAN connectivity with central processing unit [5].

The data will be collected from all the networks and analyzed offline at the central processing platform in the presence of specialists and decision makers. Here we want to collect the sensor readings from a set of points in an area over a period of time in order to detect the trends and interdependencies. The suitable application algorithms generated out of the expertise in transition modeling of draught are made use of in information processing. Then the drought pattern will be developed and evaluated by the data that is collected for several months in order to look for long term and seasonal trends.

Data is periodically transmitted from child node to parent node up the tree structure until it reaches the sink. With the tree based data collection each node is responsible for forwarding the data of all its descendents. As the binary tree structure is used it will assure that data sent is not same and its not a duplicated. Both the child nodes are not connected with each other, they are connected with its parent node further to sink node. Once the network is configured each node periodically samples its sensors and transmits its data up the routing tree and back to the base station. The typical reporting period is expected to be 30 minutes or even higher as environmental parameters like temperature humidity, pressure, wind speed, wind direction, ground water presence and precipitation will not change quickly. Coverage of the network is not equal to the range of wireless communication links being used; multihop communication techniques can extend the coverage of the network well beyond the range of radio techniques.

## 5. Data transfer to central processing unit

Source nodes communicate with its peer nodes to forward data packet to the local base station. Wireless sensor network can be divided into set of intercommunicate nodes and data flow is shown in "Fig.2," [4]. The nodes will be in sleeping state until the stimulus from the environment stimulates it. Once it gets stimulated it will become active. Sensors respond to changes in their environment by producing electrical signal. Most frequently this is change in temperature, wind pressure, and precipitation. At the sensor node, sensed data from sensors, which is in the form of analog, is converted to digital data by computational part of the node called Motes. At communication management, radio transmission is used to communicate with base station through other neighbor sensor nodes.

Wireless LAN for base station from one sensor network to base station of other sensor network. WAN connection for local base stations to remote base stations. Data from all the base station is transferred to the laboratory or office for initial processing, formulating and accumulating. Data is formulated and processed into human readable form like graphs and Tables. Finally formatted data is sent to the central processing unit for storage in database for future use. Specialists can reuse the data stored at the database at any time to run application level programs.

## 6. Increasing the coverage area

The proposed application deals with covering the very large geographical area of 10000 sq kms. Range of the sensor nodes is limited in its own way. So deploying the sensor nodes in large geographical area and coverage is a very important issue of the wireless sensor networks. Energy efficiency in wireless sensor networks is important due to very small power supply provided to it this has to last as long as possible. Energy consumption occurs in sensing, data processing and communications. A major concern in wireless sensor networks is to maximize the coverage area while maximizing the life span of the wireless sensor networks. A point is said to be physically covered if the point is within the sensing distance of at least one of the sensor node and at the notion of information coverage ,sensor nodes makes a decision for a data to sense at particular location [7].

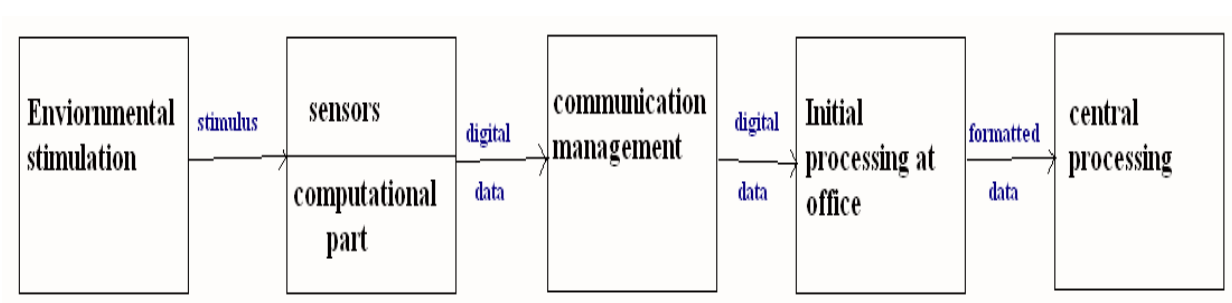

Fig.2. Block diagram of data flow

Fig.3. Hexagonal model

### 6.1 The cell shape

The cellular region should be as large as possible to reduce the number of nodes [6]. The coverage problem in wireless sensor networks is to place a sensor network in a communication area so that entire service area is observed or sensed. A network with better coverage provides better quality of service since it will be able to monitor its area of phenomenon more effectively. Shape of cell is chosen so that it should cover as large area as possible. The common available shapes for cell are square, circle, equilateral triangle, hexagon, and other polygons. With large coverage area of the cell will have the advantages like reducing flooding packets , saving a lot of power of the network in turn increasing the lifespan of the network, saves the bandwidth of the network and more importantly it reduces the no of nodes required by the network leading to the better cost performance.

Each cell covers certain area which is called as footprint and they are determined by measurements of the fields. Systematic design is done with the help of regular cell shape and adaptation of future growth .its quite natural to choose the cell shape as circle to represent the coverage area of the base station [8].With the circle as the Coverage area when applied to the sensor network each node will cover area of 10.18 sq kms approximately. While the circle cannot be overlaid upon map without leaving the gaps or overlap with each other and now taking the geometric regions which will not form overlap or wont leave gaps are square, equilateral triangle, regular hexagon.

Hexagonal shell shape is a conceptual model of the radio coverage of the node shown in "Fig.3,". It will serve the weakest radio terminals within the footprint [8]. With the hexagonal model we can cover the area of approximately 11.17 sq kms. With an incremental in 1 km coverage as compared with the circle model. With large area coverage we can achieve less number of the nodes due to large coverage by each node.

## 7. Software platform for the system

There are two types of application programs that are run at the base stations.

Network application program.

Drought prediction application program

At the network application program sensor nodes coordinate between themselves for data collection at the local base station. The integration of data at the base station is carried out by wireless sensor network protocols.

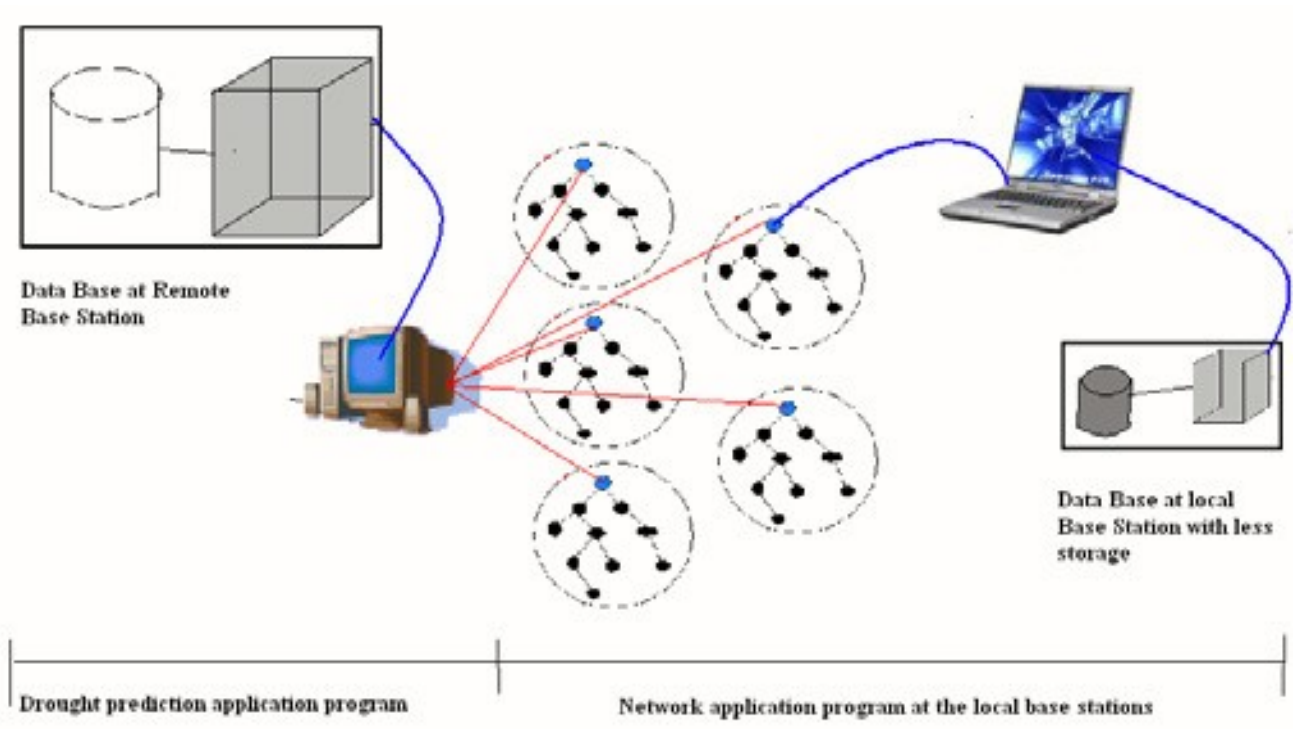

Fig.4. Database management

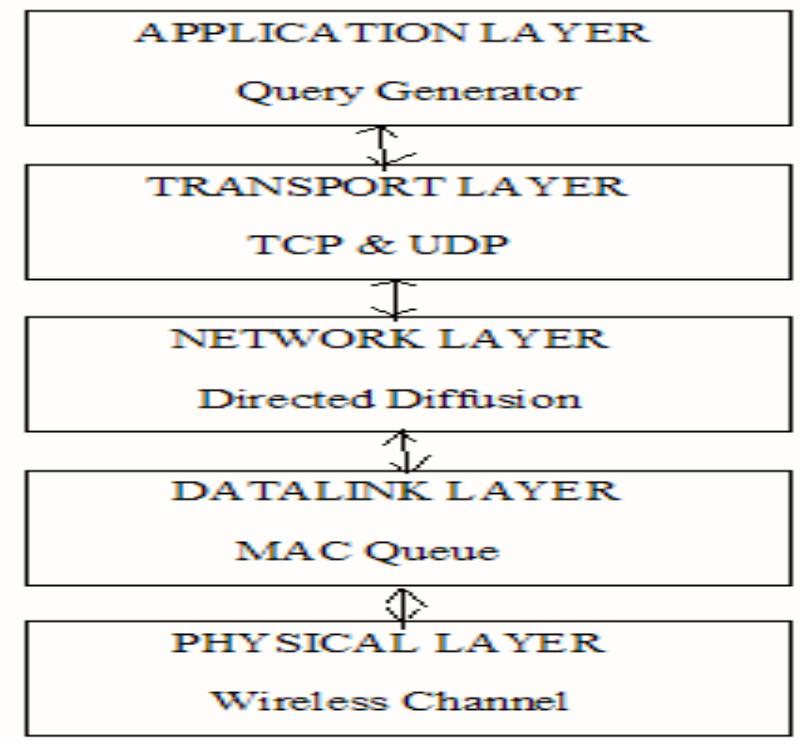

Fig.5. communication architecture

The local base stations at local sensor networks are provided with database of low storage capabilities as shown in "Fig.4". All the data collected by sensor nodes is stored at the local base stations. The database stores time stamped readings from the sensors, health status of individual sensors, connectivity and routing as well as sensor locations. It will record both raw and calibrated sensor readings in addition to above attributes. Routing layer is to be installed to get the data from node to local base station. TinyOS is an open-source operating system, which is a component-based operating system that enables easy linking of extra functionality into a program and specific parts can be replaced with other implementations. It features an event-driven execution model and has no support for threads. A program needs to implement the event handlers for all events (interrupts) it wants to handle [13].

Network applications are made to run at the local base stations. Data visualization software makes easy to connect database, to analyze, and to graph sensor readings like temperature, precipitation, pressure and humidity. The data visualization software provides software for viewing and manipulation sensor network data [12]. Drought prediction application program is run at the central data collecting point that is a remote base station as shown in "Fig.4,". The data integrated at local base station is transmitted to the remote base station or central data collecting point, establishing a wide area network connection between local base stations and the central data collection point. A comprehensive observational database of wide geographical region produced by the distributed sensor networks is managed here. It has got 24X7 active communication manager collecting information. The drought prediction application program is developed using high-level languages and libraries. By applying various principles of identifying, tracking and estimating sensor data trends prediction of drought is possible by developing drought evolution pattern.

## 8. Total communication model

In our application sensor nodes are constrained in energy supply thus techniques to eliminate energy inefficiencies that shorten the lifetime of the network [3] is applied. In these wireless sensor networks nodes do not have global IDs like IP addresses as in many other networks. So for accessing a node any protocol can use the attribute based naming and location based addressing [3]. For attribute value pairs we will be more interested in querying attribute of phenomenon rather than querying an individual node [3].

The data communication between two nodes will take place in layer by layer fashion as shown in "Fig.5,". Layers will communicate with each other by gates. Each layer will have gates to other layers of the node. i.e. data from upper layer is sent to lower layer through layer protocol and vice versa [14]. The top most layer application layer communicates with transport layer through gates. In this layer queries are generated by query generator which may be of attribute value or data rate [14]. These queries are tasked to start collecting data. Incase of need, the system administrators is enabled to interact with sensor network for querying the data from the nodes that can be done by sensor management protocol (SMP) as said in [3]. If the user needs the data collected by the nodes he can send his interest to sensor node asking about certain readings of precipitation, wind pressure, and temperature. This can be achieved by the task assignment and data advertisement protocol as said in [3].

The designed system covers a very vast area for predicting the severity of drought. The transport layer will communicate with application layer and network layer to co-ordinate the communication [14]. The system is accessed through internet for analysis and taking decision by the specialists, for that the transport layer protocols are used mainly TCP and UDP. Since the system has to work for the most part of the year energy is the main concern in the network. Each sensor nodes will have very limited amount of memory too, hence for communication between user and the sink nodes is done by TCP or UDP, but the communication between the sensors nodes and sink nodes will be by UDP [3] so as to save energy in communication.

In Network layer the query generated by application layer passed by the transport is handled and sent to lower layer by Protocols like flooding, each packet is broadcasted to all. It moves until a node receives a data with a maximum no of hops to be reached or node itself is the destination of packet [9]. Here the directed diffusion protocol is used to transfer data from node to node. Directed diffusion is data centric in which communication is for named data. Directed diffusion can achieve energy savings by selecting good paths, by caching and data processing in the network [15]. The main idea of the Data Centric paradigm is to combine the data coming from different sources and route by eliminating redundancy, minimizing the number of transmissions, thus saving network energy and prolonging its lifetime.

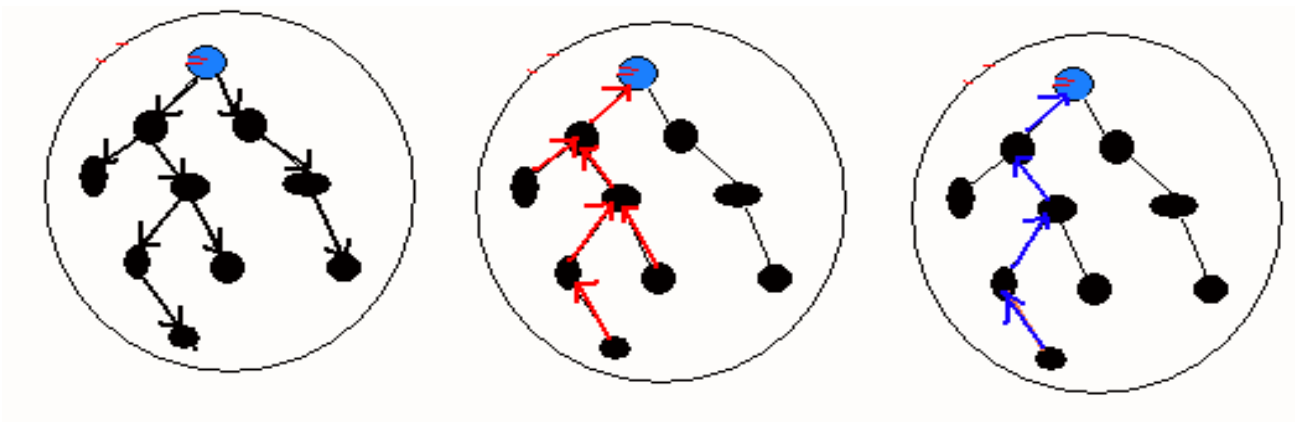

Fig.6. Directed diffusion (from left, propagate interest, setup gradient, send data)

In directed diffusion, source node requests data by sending interests and data that matches the interest is drawn towards the node with intermediate nodes caching and directing the data towards the source node. Messages exchanged between nodes are within some vicinity [15]. Sensors measure events and create gradients of information in their respective neighborhoods as shown in "Fig.6,". The sink requests data by broadcasting interests. An interest describes a task required to be done by the network. An interest diffuses through the network hop by hop, and is broadcast by each node to its neighbors. As the interest is propagated throughout the network, gradients are set up to draw data satisfying the query toward the requesting node [10].

In directed diffusion sink sends interest as it is based on the data centric routing, an interest is an interrogation that asks for user requirement supported by sensor network and it is task description to all sensors of the network as shown in the "Fig.6," as said in [3]. The task descriptors are named by assigning attribute value pairs that describe the task. Each sensor node will store the interest entry in its cache. As the interest propagates through sensor field the gradients from the source back to the sink are set up. As shown in "Fig.6 (b),". Gradient is the direction state created in each node that receives the interest and it is towards its immediate predecessor which sent interest to that node. When the source has the data for the interest, source sends the data and it starts flowing towards its originators of interest along the interest's gradients path [15]. The interest, data propagation and aggregation are determined locally. Sink will refresh and reinforce the interest when it starts to receive data from the source [3].

In data link layer the data in form of packets from network layer are received and they are placed in wireless channel .packet contains the information related to its transmission i.e., unicast or broadcast. Here the packets are placed in the queue by MAC layer. and if the network packet length exceeds the MAC frame then it is fragmented and queued for transmission. MAC layer checks the channel for idle state for transmitting packets from MAC queue. Once the channel becomes free the packet is transmitted [14].

All the connections between nodes are controlled by wireless channel. Nodes exchange data and communicate with other node with the help of these connections. Transmission of messages takes place with some amount of delay. After data reaches its destination Data will move from bottom layer to top layer fashion [14].

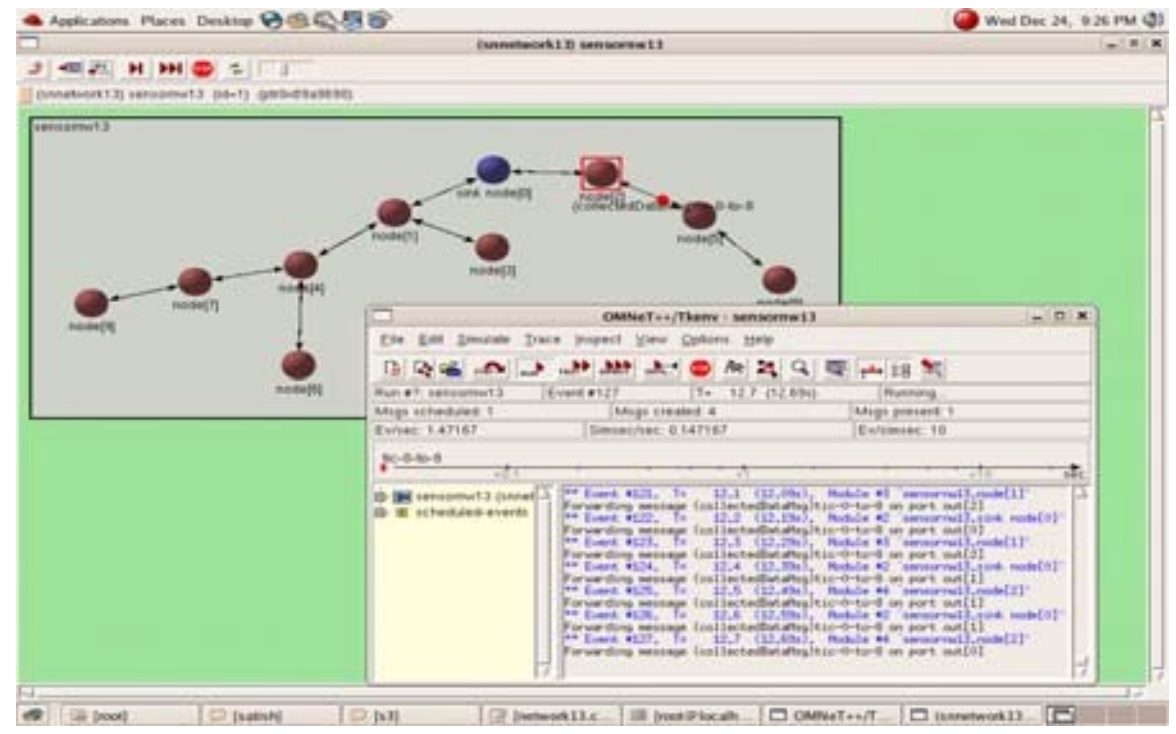

Fig.7. Snapshot of simulated network

## 9. Simulation Environment

The application is implemented with Objective Modular Network Testbed in C++ ( OMNeT++). It is a discrete event simulator and it is component based, modular simulation framework. It consist of hierarchically nested modules in which the top level module is called as system module which may further contain one or more sub modules. Modules can be nested and they are distinguished by compound or simple. Simple modules are associated C++ file and implemented using OMNeT++ class library. Compound modules are not associated with C++ files. Communication between modules is by exchange of messages. It will follow its local simulation time and it advances whenever module receives messages. Messages are scheduled by self messages. And initialization files are used to configure simulation executions [15] [16].

The topology of the model is specified using NED language as it facilitates the modular description of the network. Network description may consist of component description such as the channels, simple modules compound modules. All files containing network description will have .ned suffix. To implement the functionality of the simple modules C++ file is written. They registered in OMNeT++ with Define_Module macro. Make file is created to compile and link the program .finally omnetpp.ini is created, which tells the simulation program which network to simulate [17].

In our application the network used for simulation is binary tree network in which the child nodes will be connected to the parent node and further to the sink. In our network each network will contain 10 nodes including sink. Sink will act as the root node and the communication from leaf node to the sink node is done through in between nodes of the path. Total of 50 nodes form the full sensor network for drought prediction. Communication of messages between modules will be done with some amount of delay as shown in "Fig.7,". Simulation time will follow the local simulation time of the network and it depends on the no of hops the message takes to reach the sink from the source node. Temperature, precipitation, pressure, humidity are used as parameters. Graphs are generated using the scalars and ploves.

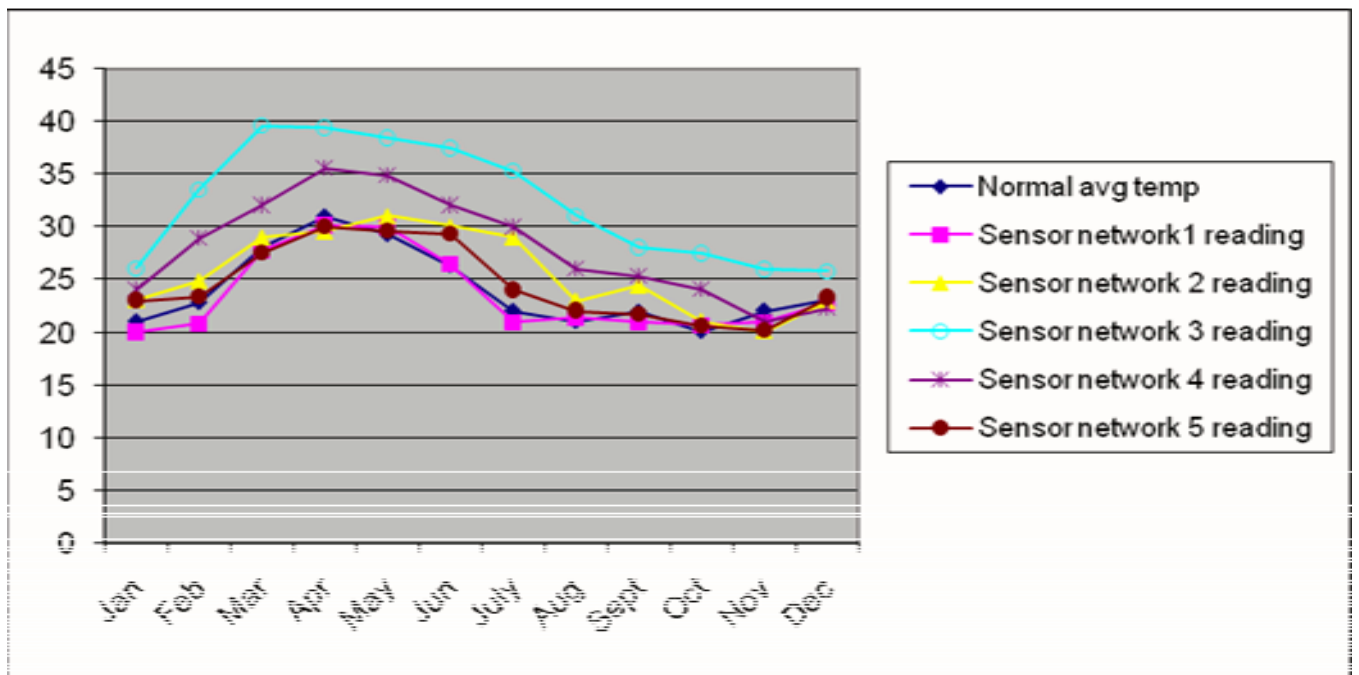

Fig.8a. Drought evolution pattern w.r.t. temperature.

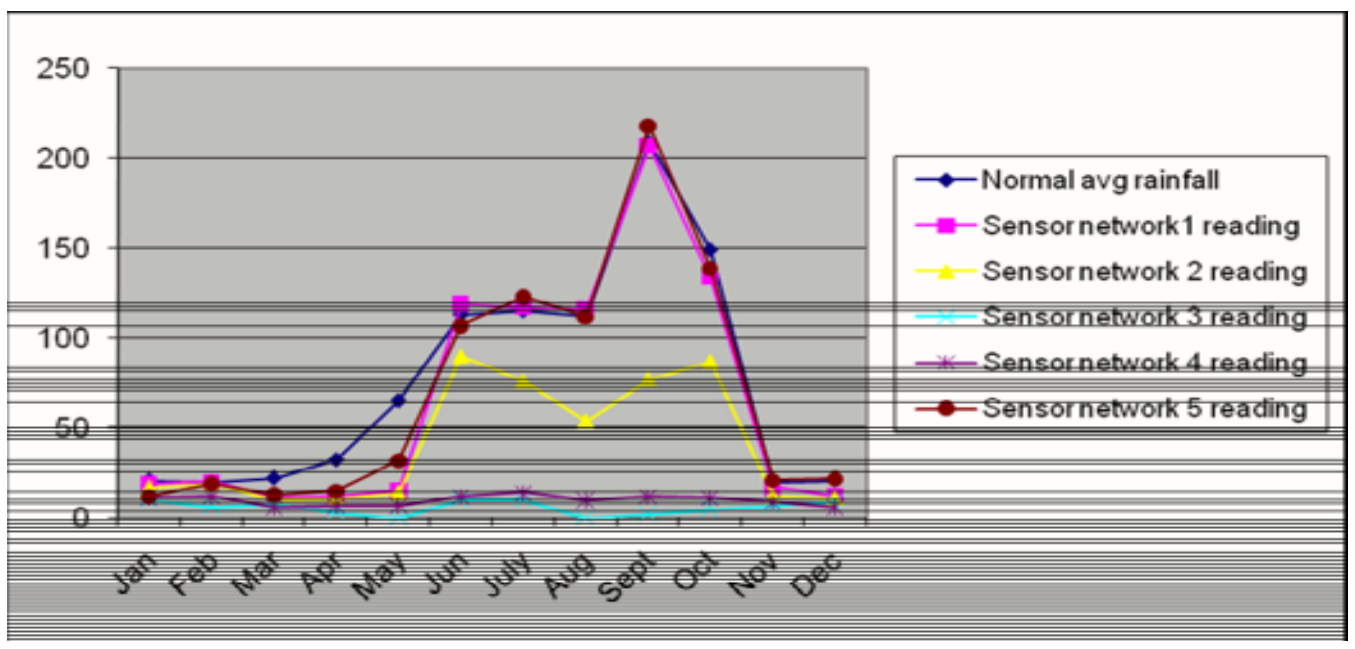

Fig.8b. Drought evolution pattern w.r.t. precipitation

## 10. Results

Data is collected from all the sensor networks and stored in the server at the central base station. Data is collected for large period of the year and from the collected data drought pattern is evolved by the specialists. In our application data collected for a year is made use of to predict the severity of drought. Drought severity is classified into 4 types and they are non drought, slight drought, moderate drought, and serious drought [17].drought evolution pattern is shown in "Fig.8,".

The drought occurs mainly due to lack of precipitation and due to high temperature. From Fig.8a.&8b. Sensor network 3 is showing severe drought as the temperature for the year remains high throughout the year and it records almost null rainfall in average throughout the year, as a result of this, sunny weather is dragged for long days and results in severe drought. Sensor network 4 is reported to suffer from moderate drought as the temperature is much above the average normal temperature and rainfall is less than 25mm approximately. Sensor network 2 is reported to be in slightly drought prone area and sensor network 1and 5 are reported to be in non drought area. With the wind pressure and wind direction data, drought is moving towards region of sensor network 4 from sensor network 3 as the wind will carry dry weather with them. It is predicted that sensor network 2 and 3 may suffer from severe drought in future. The information on drought severity is given to meteorologists to take further preventive measures.

## 11. Conclusion

This system describes the drought severity prediction scheme using wireless sensor networks. Frontier technologies and approaches are used to solve a practical problem of adverse natural phenomenon in large area. The sensor network architecture of the system and its working procedures are presented for the drought prediction purpose. Extending this model to a powerful global network for geographical and weather pattern observation could be the next step. Developing applications using this can correlate and predict natural calamities ranging from earthquakes to hurricanes that are normally occurring in the random fashion but may have cross correlation to observable physical parametric variations around the globe. This will be of great advantage.